\UseRawInputEncoding
\documentclass[aps,floatfix,prb,twocolumn]{revtex4-1}
\usepackage{graphicx}
\usepackage{dcolumn}
\usepackage{bm}
\usepackage{float}
\usepackage{placeins}
\usepackage{amsmath}
\usepackage{romannum}
\usepackage{csquotes}
\usepackage{hyperref}
\usepackage{subfigure}
\usepackage{booktabs}
\usepackage{multirow}
\hypersetup{
    colorlinks = true,
    citecolor = blue,
    linkcolor = blue,
    urlcolor = blue,
}

\begin{document}

\title{Ultra-low lattice thermal conductivity induces high-performance thermoelectricity in Janus group-VIA binary monolayers}
\author{Shao-Bo Chen$^{1,2}$}
\author{San-Dong Guo$^{3}$}
\author{Bing Lv$^{4}$}
\author{Mei Xu$^{4,*}$ }
\author{Xiang-Rong Chen$^{1,*}$}
\author{Hua-Yun Geng$^5$}
\affiliation{$^1$ College of Physics, Institute of Atomic and Molecular Physics, Sichuan University, Chengdu 610065, People’s Republic of China}
\affiliation{$^2$ College of Electronic and Information Engineering, Anshun University, Anshun 561000, People’s Republic of China}
\affiliation{$^3$ School of Electronic Engineering, Xi’an University of Posts and Telecommunications, Xi’an 710121, People’s Republic of China}
\affiliation{$^4$ School of Physics and Electronic Science, Guizhou Normal University, Guiyang 550025, People’s Republic of China}
\affiliation{$^5$ National Key Laboratory for Shock Wave and Detonation Physics Research, Institute of Fluid Physics, CAEP, Mianyang 621900, People’s Republic of China\\
\text{\bf{Email:}} {xumei6401@tom.com; xrchen@scu.edu.cn}}

\begin{abstract}
In this paper, the electrical transport, thermal transport, and thermoelectric properties of three new Janus STe$_{2}$, SeTe$_{2}$, and Se$_{2}$Te monolayers are systematically studied by first-principles calculations, as well as the comparative with available literature’s results using different methods. It is found that the Seebeck coefficient and conductivity have opposite dependence on temperature, and we illustrate this phenomenon in detail. The decrease of the thermoelectric power factor (\textit{PF}) with temperature originates from the decrease in conductivity. To obtain accurate and convergent lattice thermal conductivity, the root mean square (RMS) is calculated to obtain a reasonable cutoff radius for the calculation of third-order forces. Janus STe$_{2}$, SeTe$_{2}$, and Se$_{2}$Te monolayers exhibit ultra-low lattice thermal conductivity of 0.2, 0.133, and 4.81$\times10^{-4}$ W/mK at 300 K, which result from the strong coupling effect between the acoustic mode and the low-frequency optical branch, low phonon group velocity, small phonon lifetime, and large anharmonicity. Consequently, ultra-high $\textit{ZT}$ values of 2.11 (2.09), 3.28 (4.24), and 3.40 (6.51) for n-type(p-type) carrier doping of STe$_{2}$, SeTe$_{2}$, and Se$_{2}$Te are obtained, indicating that they are promising thermoelectric materials. 
\end{abstract}
\maketitle
\pagenumbering{arabic}

\section{Introduction}
Thermoelectric materials have become a hot research topic because they can directly convert heat energy into electric energy. As a new member of the thermoelectric materials family, two-dimensional (2D) materials have attracted extensive attention because of their unique electrical, thermal, and mechanical properties. In the past decades, the 2D thermoelectric materials, such as SnSe, Bi$_{2}$Te$_{3}$, and MoS$_{2}$, have been theoretically predicted and the corresponding samples have been experimentally synthesized \cite{RN2172, RN2171, RN2148}. These 2D materials exhibit excellent thermoelectric properties and have great potential when they are used to manufacture high-performance thermoelectric devices. However, due to the high thermal conductivity of 2D materials, the thermoelectric properties of 2D materials are still much lower than those of traditional bulk thermoelectric materials, which limits the thermoelectric applications in 2D materials.\\
Since Li et al. \cite{RN2175, RN1964} successfully synthesized 2D tellurium (Te), its attractive characteristics including thickness-dependent band gap, environmental stability, piezoelectric effect, thermoelectric effect, high carrier mobility, and light response show great potential in photodetectors, field effect transistors, piezoelectric devices, thermoelectric devices, modulators, and energy harvesting devices \cite{RN2690}. 2D allotrope tellurene composed of the metal-like element Te mainly includes three phases ($\alpha$, $\beta$, and $\gamma$). The stable $\alpha$-Te has the structure of 1T phase MoS$_{2}$, and the metastable $\beta$-Te and $\gamma$-Te have the structure of the foursquare structure and 2H phase MoS$_{2}$, respectively. Tellurene is a well-known p-type semiconductor with a band gap of 0.35 eV at room temperature. Subsequently, 2D homogeneous selenene ($\alpha$-Se, $\beta$-Se, $\gamma$-Se) with the same structure was successively discovered \cite{RN2033}. The 2D material family is enriched again by the appearance of tellurene and selenene. Among these 2D materials composed of a single element, tellurene and selenene have the lowest lattice thermal conductivity among 2D materials reported so far \cite{RN2033, RN2036, RN2040}. It has been theoretically and experimentally demonstrated that compounds composed of both Te and Se elements have excellent thermoelectric and electron transport properties \cite{RN2278, RN2267}. Inspired by the noncentrosymmetric Janus structure have novel properties and group-VIA elements possess ultra-low lattice thermal conductivity, it is envisaged that the Janus-structured binary compounds composed of group-VIA elements (S, Se, Te) have excellent thermoelectricity as a consequence of ultra-low lattice thermal conductivity, which has potential applications in thermoelectric devices. In this paper, we predicte three stable group-VIA binary monolayers with ultra-low lattice thermal conductivity and ultra-high thermoelectric performance.

\section{Calculation details}\label{sec:Calculation}
All calculations based on the density functional theory are performed in the VASP code \cite{RN1057}. The projector-augmented-wave (PAW) pseudopotential is carried out to represent the interaction between the ions and the electrons \cite{RN2019}. The generalized gradient approximation (GGA) within the Perdew-Burke-Ernzerhof (PBE) formulation is employed for the exchange-correlation potential. The second-order force constants (harmonics) are implemented in the PHONOPY code \cite{RN1862} based on density functional perturbation theory (DFPT) using a 5 $\times$ 5 $\times$ 1 supercells and 3 $\times$ 3 $\times$ 1 k-point grid. Then, the phonon dispersion relation can be obtained by diagonalizing the force constant matrix. The ab initio molecular dynamics (AIMD) simulations are performed by using the canonical ensemble with a 5 $\times$ 5 $\times$ 1 supercell and the simulation time is set to 5 ps with the step of 2 fs. Based on Boltzmann transport theory, TransOpt \cite{RN2085} is used to calculate the electrical transport coefficient of the material with constant electron-phonon coupling approximation in the relaxation time approximation method. The third-order force constants (non-harmonics) are calculated using the finite displacement method \cite{RN2241, RN1862} with a 4 $\times$ 4 $\times$ 1 supercell and a 3 $\times$ 3 $\times$ 1 k-point grid. To get the precise lattice thermal conductivity, we separately test the nearest-neighbor interactions and the Q-grid. The lattice thermal conductivity is obtained by iteratively solving the phonon Boltzmann transport equation using the ShengBTE code \cite{RN1968}.
\section{Result and discussion}
\label{sec:Result}

\subsection{Electrical transport properties}\label{sec:Electrical}
\begin{figure}
	\centering
	\includegraphics[width=1.0\linewidth]{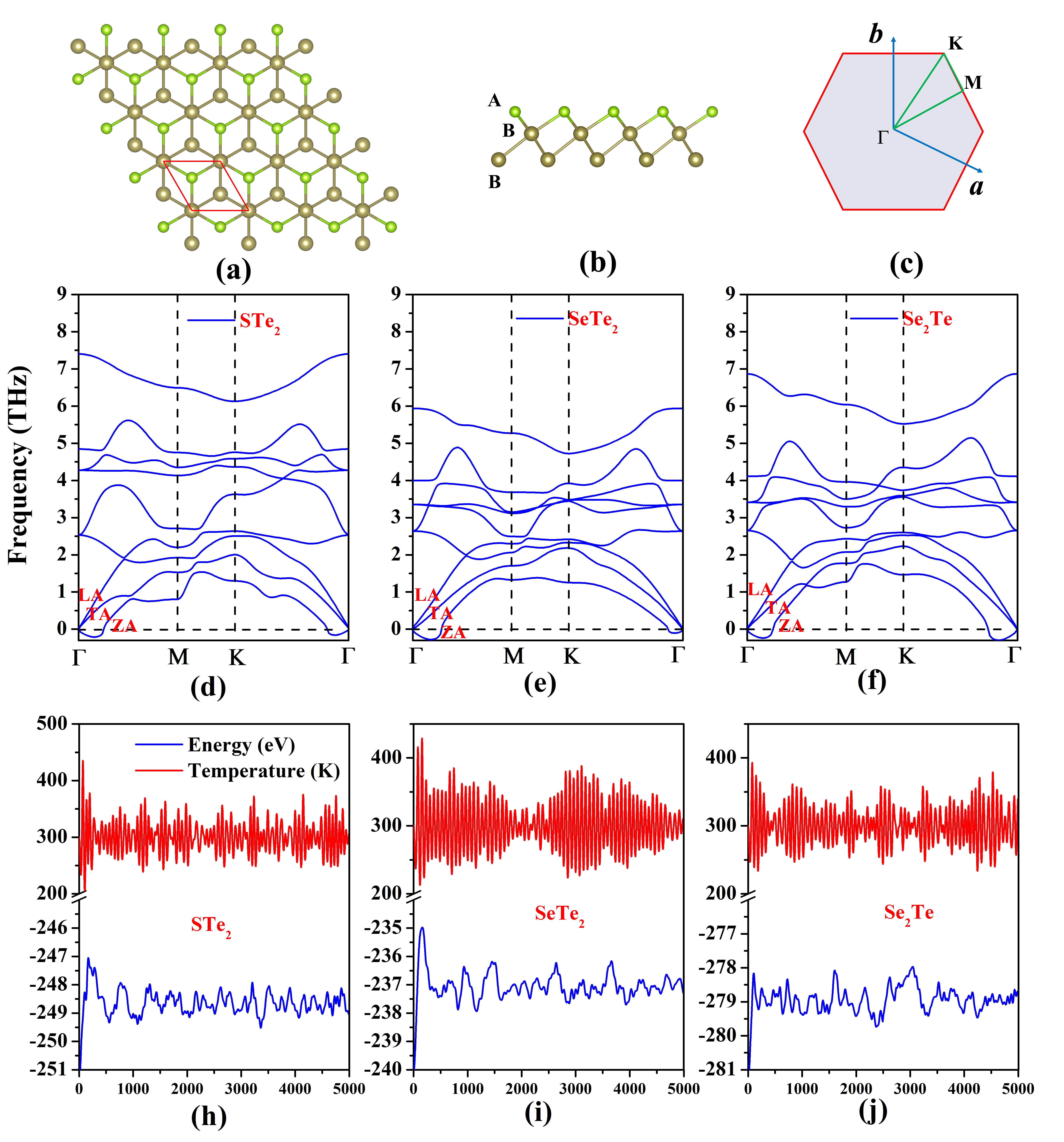}
	\caption{(Color online) (a) and (b) represent the top and side view of Janus structures, respectively, together with the (c) first Brillouin zone high symmetric path. The phonon dispersions of (d) STe$_{2}$, (e) SeTe$_{2}$, (f)Se$_{2}$Te monolayers. The three low-frequency acoustic branches are divided into one out-of-plane vibration acoustic branch (ZA) and two in-plane transverse (TA) and longitudinal (LA) vibration acoustic branches. (h-j) represents the AIMD simulation: the variation of the total energy (blue) and temperature (red) with time.}
	\label{fig:fig1}
\end{figure}

There are six different VIA binary monolayers, depending on the combination of different elements. Only three structures are stable after the stability test. The Janus STe$_2$, SeTe$_2$, and Se$_2$Te monolayers belong to P3m1 space group with $\textit{C}_{3v}$ point symmetry, as shown in Fig. \ref{fig:fig1}(a, b). The optimized lattice constants are 4.03, 4.11, and 3.88 \AA \quad for STe$_2$, SeTe$_2$, and Se$_2$Te monolayers, respectively, which are consistent with the ones in the literatures \cite{RN1880, RN2563, RN2560, RN2722}. To check the stability of these systems, in Fig. \ref{fig:fig1}(d, e, f), we calculate the phonon dispersion along with a high symmetric path ($\Gamma$-M-K-$\Gamma$, see Fig. \ref{fig:fig1}(c)) in the Brillouin zone, and the frequency of phonon is free of imaginary frequency, showing that these systems are dynamic stability, which can also be confirmed in the previous reports \cite{RN2560, RN2722}. In addition, the AIMD simulations show that these systems are thermally stable as a consequence of the energy and temperature always vibrate in small amplitude around a fixed value, as shown in Fig.\ref{fig:fig1}(h, i, g). To further determine the energy stability of the system, we calculate the cohesive energy according to the formula in the references \cite{RN2724}, as listed in Table \ref{tab1}, and the results are compared with those in other references \cite{RN2560,RN2722}. The cohesive energy of all negative values indicates that the systems are energetically stable. Then, we first calculate the band structures of those Janus monolayers. The spin-orbit coupling (SOC) effect is considered in the calculation of the band structures because of the heavy element Te in the system. The results show  that the STe$_2$, SeTe$_2$, and Se$_2$Te monolayers are all indirect band-gap semiconductors with a band-gap of 0.74(1.23), 0.53(0.98), and 0.67(1.06) eV at PBE+SOC (HSE06+SOC) methods, respectively, which agrees well with the reports \cite{RN2560,RN2722}.\\
Before using the TransOpt code \cite{RN2085} to calculate the electronic transport properties, several important parameters that need to be prepared are the electron (hole) deformation potential $\textit{E}_{l}$, Young's modulus $\textit{G}$, and the Fermi level $\textit{E}_{f}$. The detailed input parameters are shown in Table \ref{tab1}. $\textit{C}_{2D}$ is the 2D elastic modulus, which is calculated using the formula $C_{2 D}=\frac{1}{S_0} \frac{\partial^2 E}{\partial\left(\Delta l / l_0\right)^2}$. By comparison, our calculated results are closer to the references \cite{RN2560}. \textit{E} is the total energy of the material under a small uniaxial strain ranging from $-2$\% to +2\% with a step of 0.5\%. $\textit{l}_{0}$ is the lattice constant of the equilibrium structure, $\Delta$\textit{l} is the change in lattice constant. $\textit{E}_{l}$ is the deformation potential in the transport direction of the valence band maximum (VBM) and conduction band minimum (CBM), and its expression is $E_l=\frac{\partial E_{\text {edge }}}{\partial\left(\Delta l / l_{0}\right)}$, $\textit{E}_{edge}$ is the energy of VBM or CBM under small uniaxial strain. Through Table \ref{tab1}, it is easy to find that our calculation results are consistent with the previous calculation results, indicating the correctness of the calculation.
\begin{table}[htbp]
	\centering
    \setlength\tabcolsep{4pt}
    \renewcommand\arraystretch{1.3} 
	\caption{ The summarized elastic modulus $\textit{C}_{2D}$ of the 2D material, the deformation potentials $\textit{E}_{l}^{VBM}$ and $\textit{E}_{l}^{CBM}$ near the VBM and CBM, as well as Young's modulus $\textit{G}$ and the Fermi level $\textit{E}_{f}$.}	
\begin{tabular}{cccccc}
		\hline
		\hline
		Compounds & $\textit{C}_{2D}$& $\textit{E}_{l}^{VBM}$ & $\textit{E}_{l}^{CBM}$& $\textit{G}$& $\textit{E}_{f}$\\
		&(N/m)&(eV)&(eV)&(GPa)&(eV/atom)\\
		\hline
		Janus STe$_2$ & 41.75  & $-4.34$  &$-6.27$ & 116.57  & $-1.69$ \\
		              &40.50 \cite{RN2560}& &  &  &$-3.35$\cite{RN2560}\\
		              &50.91 \cite{RN2722}& 3.92\cite{RN2722} &  6.06\cite{RN2722} &  & $-2.38$\cite{RN2722}\\
		Janus SeTe$_2$ & 38.35  & $-4.78$   & $-6.42$ & 104.57  & $-2.00$ \\
	                	&46.49 \cite{RN2722}&4.36\cite{RN2722}  & 6.18\cite{RN2722} &  &$-2.24$\cite{RN2722}\\
		Janus Se$_2$Te & 42.9  &$-7.77$  & $-9.72$& 124.97  & $-2.19$ \\
		\hline
		\hline
\end{tabular}
	\label{tab1}
\end{table}

\begin{figure}
	\centering
	\includegraphics[width=1.0\linewidth]{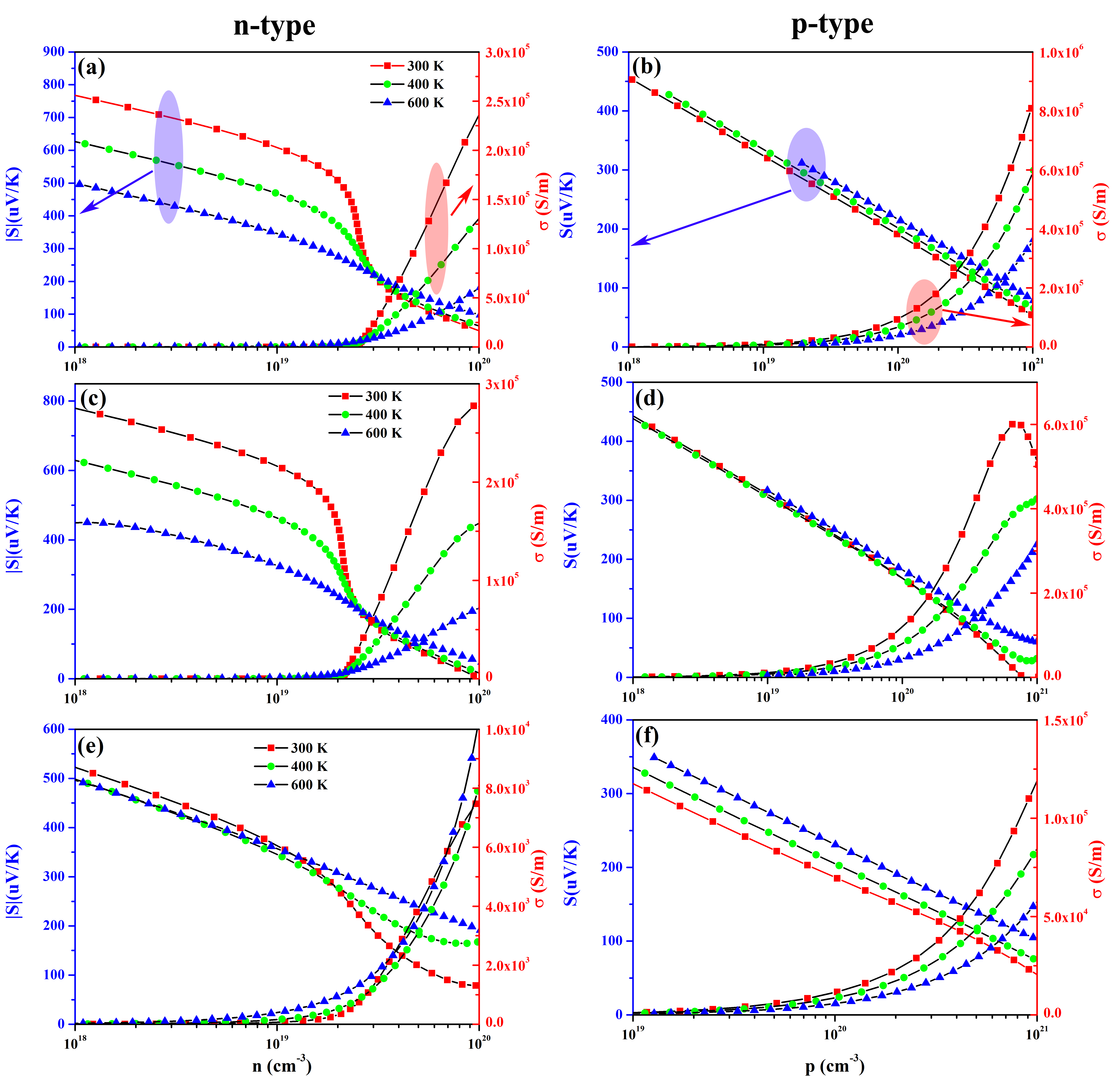}
	\caption{(Color online) Seebeck coefficient S and conductivity $\sigma$ of (a, b)STe$_{2}$, (c, d)SeTe$_{2}$ and (e, f)Se$_{2}$Te as a function of doping concentration at different temperatures.  The light blue and light red areas in the figure help to distinguish between the $\textit{S}$ and $\sigma$ corresponding to the ordinate.}
	\label{fig:fig2}
\end{figure}

\begin{figure}
	\centering
	\includegraphics[width=1.0\linewidth]{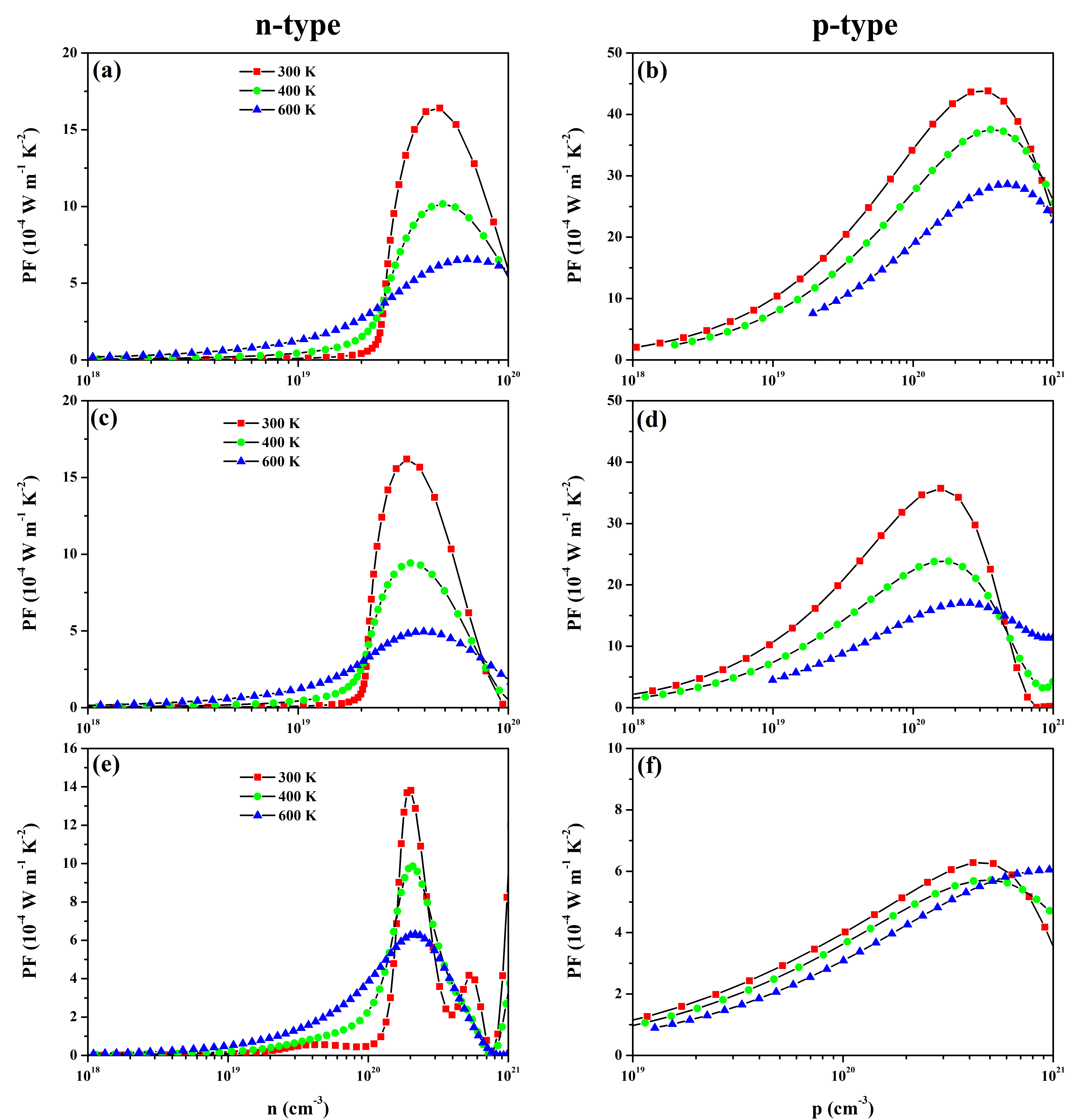}
	\caption{(Color online)  Power factor \textit{PF} of (a, b) STe$_{2}$, (c, d) SeTe$_{2}$ and (e, f) Se$_{2}$Te as a function of doping concentration at different temperatures.}
	\label{fig:fig3}
\end{figure}

Fig.\ref{fig:fig2} depicts the carrier-concentration-dependent Seebeck coefficient \textit{S} and conductivity $\sigma$ of Janus binary compounds. For both n-type doping and p-type doping, \textit{S} monotonically decreases with the increase of carrier concentration, while $\sigma$ exponentially increases with the increase of carrier concentration. The power factor \textit{PF}, $PF=S^2\sigma$, hinder as a result of the strong coupling effect between \textit{S} and $\sigma$, thereby limiting the thermoelectric figure of merit $\textit{ZT}$. Therefore, the decoupling between Seebeck coefficient and conductivity has become an effective strategy for improving thermoelectric efficiency. It can be observed that in the vicinity of the optimal doping concentration range (the region where \textit{S} and $\sigma$ intersect), \textit{S} tends to increase with increasing temperature. Conversely, $\sigma$ decreases with increasing temperature. This is mainly because \textit{S} is positively correlated with temperature T\cite{RN2482}:
\begin{equation}
	 S=\frac{8 \pi^{2} k_{\mathrm{B}}^{2}}{3 e h^{2}} m^{*} T\left(\frac{\pi}{3 n}\right)^{2/3}
	\label{Eq1}
\end{equation}
where $\textit{m}^{*}$ and \textit{n} are the effective mass of electrons (holes) and the carrier concentration, respectively. The conductivity is positively related to mobility with the relationship:
\begin{equation}
\sigma=n e \mu=n e \frac{e h^3 C_{2 D}}{k_B T m^* m_d E_l^2}
\label{Eq2}
\end{equation}
 where the $\mu$ is carrier mobility, and \textit{n} is carrier concentration. C$_{2D}$, $\textit{E}_{l}$, $k_{B}$, $\textit{m}^{*}$, and $\textit{m}_{d}$ are 2D in-plane elastic modulus, deformation potentials, Boltzmann constant, effective mass, and average effective mass, respectively. The scattering of carriers augments as a consequence of the increasing temperature, which induces a decrease in the mobility of carriers. Therefore, the conductivity is inversely related to the temperature. 
At different temperatures (300 K, 400 K, 600 K), the power factor \textit{ PF} varies with the carrier concentration, as shown in Fig.\ref{fig:fig3}. Notably, the power factor \textit{ PF} gradually decreases with the increase of temperature. With a combination of temperature-dependent conductivity and the Seebeck coefficient discussed above, one can obtain that the decrease of \textit{PF} with temperature is mainly caused by the decrease of conductivity with temperature.   At 300 K, the maximum power factor of those systems are summarized in Table \ref{tab2}, together with other results. The maximum power factor $\textit{PF}_{max}$ of STe$_{2}$ and SeTe$_{2}$ is larger than that of Se$_{2}$Te. For example, the n-type (p-type) PFmax of STe$_{2}$ and SeTe$_{2}$ are 16.41 (43.81)$\times10^{-4}$ W/mK$^{2}$ and 16.20 (35.73)$\times10^{-4}$ W/mK$^{2}$ is larger than 13.81 (6.28) $\times10^{-4}$ W/mK$^{2}$ of Se$_{2}$Te, which is may mainly because the first two materials contain two Te elements with optimal thermoelectric properties. The $\textit{PF}_{max}$ of these three materials is comparable to those of typical excellent thermoelectric materials such as SnSe (∼4$\times10^{-3}$W/mK$^{2}$ at 300 K), Bi$_{2}$Te$_{3}$ ($\sim$3.5$\times10^{-3}$ W/mK$^{2}$ at 300 K), PbTe ($\sim$2.5$\times10^{-3}$ W/mK$^{2}$ at 500 K), 1T''-phase MoSe$_{2}$ ($\sim$6$\times10^{-3}$W/mK$^{2}$ at 200-500 K), and doped MoS$_{2}$ (2.98$\times10^{-3}$ W/mK$^{2}$) \cite{RN2608, RN2610, RN2607, RN2606, RN2609}, indicating that STe$_{2}$, SeTe$_{2}$, and Se$_{2}$Te are high-performance 2D thermoelectric materials. \\
However, it cannot be ignored that there is a big difference between the reference’s data \cite{RN2722} and our calculation results, which may be caused by the use of different calculation parameters and calculation software. BoltzTraP, which realizes the calculation of transport coefficient based on Boltzmann transport theory, is the most widely used program by far. However, BoltzTraP adopts the approximate method of constant relaxation time, which can only calculate the ratio of conductivity, electron thermal conductivity, and other coefficients to the relaxation time. Hence, to determine the electrical transport properties, the value of relaxation time must be set by the simulation experiment or calculations. The relaxation time is closely related to the effective mass of the electron and hole,  $\tau=\frac{\mu \mathrm{m}^*}{e}$($\tau$, $\mu$, $m^{*}$ are the relaxation time, carrier mobility, and effective mass, respectively). It is well known that the calculation process of fitting the effective mass of the electron and hole with the energy band data is complicated and easy to causes errors. Thus, the dependence on the relaxation time constant introduces uncertainty in the calculation of electrical transport properties and limits the accuracy of its prediction. In contrast, the constant electro-acoustic coupling approximation method is used in the TransOpt software package, which can predict the relaxation time under phonon action and improve the accuracy of the prediction results of electrical transport properties such as conductivity. Thus, our calculations may be more accurate and reliable.

\begin{table*}[htbp]
	\centering
	\setlength\tabcolsep{8pt}
	\renewcommand\arraystretch{1.5} 
	\caption{ The calculated lattice thermal conductivity $\textit{k}_{l}$ is compared with the ones in other reports calculated by the different methods and $n^{th}$ nearest neighbors, as well as the effective thicknesses $\textit{d}$, and the n-(p-)type maximum power factor $PF_{max}$ and the maximum $\textit{ZT}_{max}$ at room temperature.}	
	\begin{tabular}{cccccc}
		\hline
		\hline
		Compounds &   &$\textit{k}_{l}$ (W/mK)&\textit{d}(\AA)& $\textit{PF}_{max}$(10$^{-4}$ W/mK$^{2}$)&\textit{ZT}$_{max}$\\
		\hline
		Janus STe$_2$ & \multirow{2}[0]{*}{this work} &0.2[9$^{th}$, ShengBTE, iteration] &\multirow{2}[0]{*}{7.12}&\multirow{2}[0]{*}{16.41(43.81)[TransOpt]}   &\multirow{2}[0]{*}{2.11(2.09)} \\
		& &$\sim$0.7[5$^{th}$, ShengBTE, iteration]&   &   &\\
		&other work\cite{RN2560}&1.16[5$^{th}$, phono3py, iteration]  & 7.18 &  &\\
		&\multirow{2}[0]{*}{other work\cite{RN2722}}&1.18[17$^{th}$, AlmaBTE, iteration] &\multirow{2}[0]{*}{7.20} & \multirow{2}[0]{*}{83(350)[BoltzTraP2]} &\multirow{2}[0]{*}{1.16(2.4)}\\
		&        &0.39[17$^{th}$, AlmaBTE, RTA]&   & &\\
		Janus SeTe$_2$ & this work &0.133[7$^{th}$, ShengBTE, iteration] &7.58& 16.20(35.73)[TransOpt]  &3.28(4.24) \\
		&\multirow{2}[0]{*}{other work\cite{RN2722}}&0.73[17$^{th}$, AlmaBTE, iteration] &\multirow{2}[0]{*}{7.79} & \multirow{2}[0]{*}{85(340)[BoltzTraP2]} &\multirow{2}[0]{*}{1.58(3.1)}\\
		&        &0.32[17$^{th}$, AlmaBTE, RTA]&   & &\\
      Janus Se$_2$Te & this work &4.81$\times$10$^{-4}$[9$^{th}$, ShengBTE, iteration] &7.39& 13.81(6.28)[TransOpt]  &3.40(6.51) \\
		\hline
		\hline
	\end{tabular}
	\label{tab2}
\end{table*}


\subsection{Thermal transport properties}
\label{sec:Thermal}

\begin{figure}
	\centering
	\includegraphics[width=1.0\linewidth]{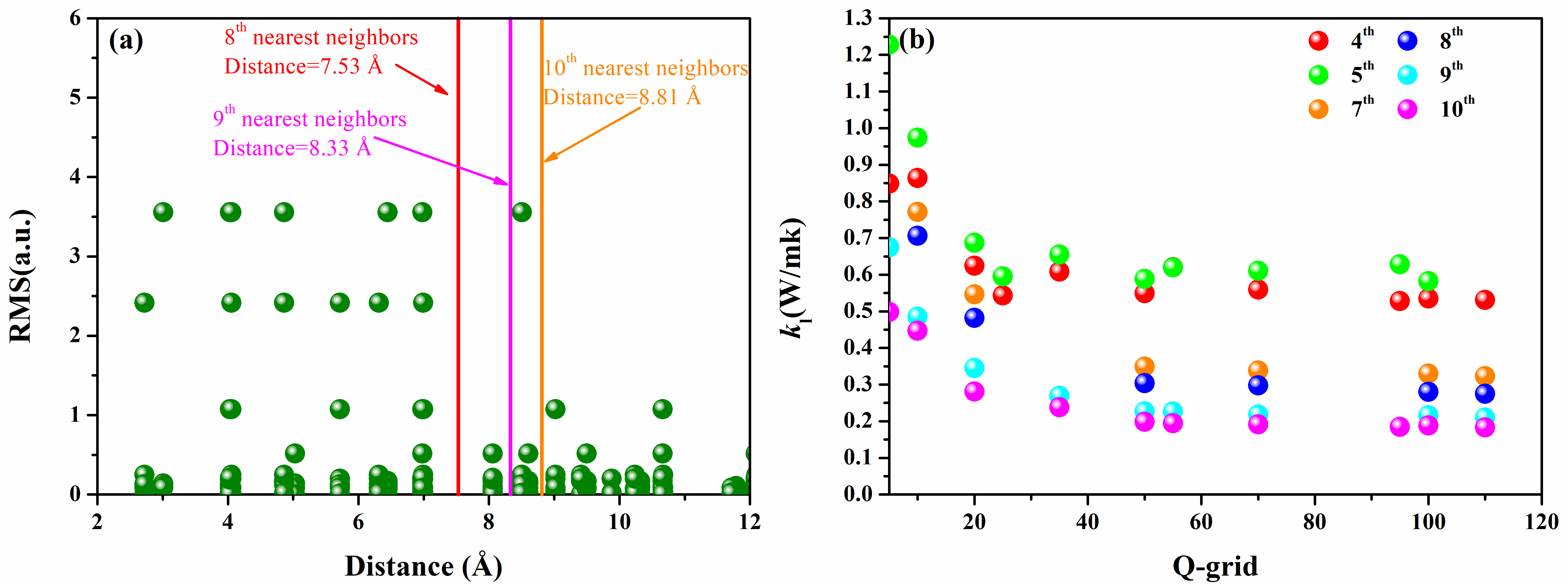}
	\caption{(a) The relationship between the RMS of STe$_{2}$ and the cutoff radius reveals the distant interaction between atoms. (b) Convergence test of lattice thermal conductivity $\textit{k}_{l}$ of STe$_{2}$ with the change of Q grid when considering different cutoff radius. }
	\label{fig:fig4}
\end{figure}

Currently, a very effective method to obtain lattice thermal conductivity is combining first-principles calculations based on an anharmonic lattice dynamics method with the phonon Boltzmann transport equation. This method involves the calculation of interatomic force constants (IFCs) with the harmonic IFCs and the higher-order IFCs \cite{RN2688, RN2689}. To obtain satisfactory lattice thermal conductivity $\textit{k}_{l}$, the convergence of the cutoff radius for anharmonic interatomic force constants (IFCs) is tested. However, it is not easy to evaluate higher-order IFCs in comparison to harmonic IFCs. Hence, we use the root mean square (RMS) \cite{RN2573, RN2571} to evaluate the strength of interatomic interactions, which consist of the elements of the harmonic IFC tensor, that is 
\begin{figure}
	\centering
	\includegraphics[width=1.0\linewidth]{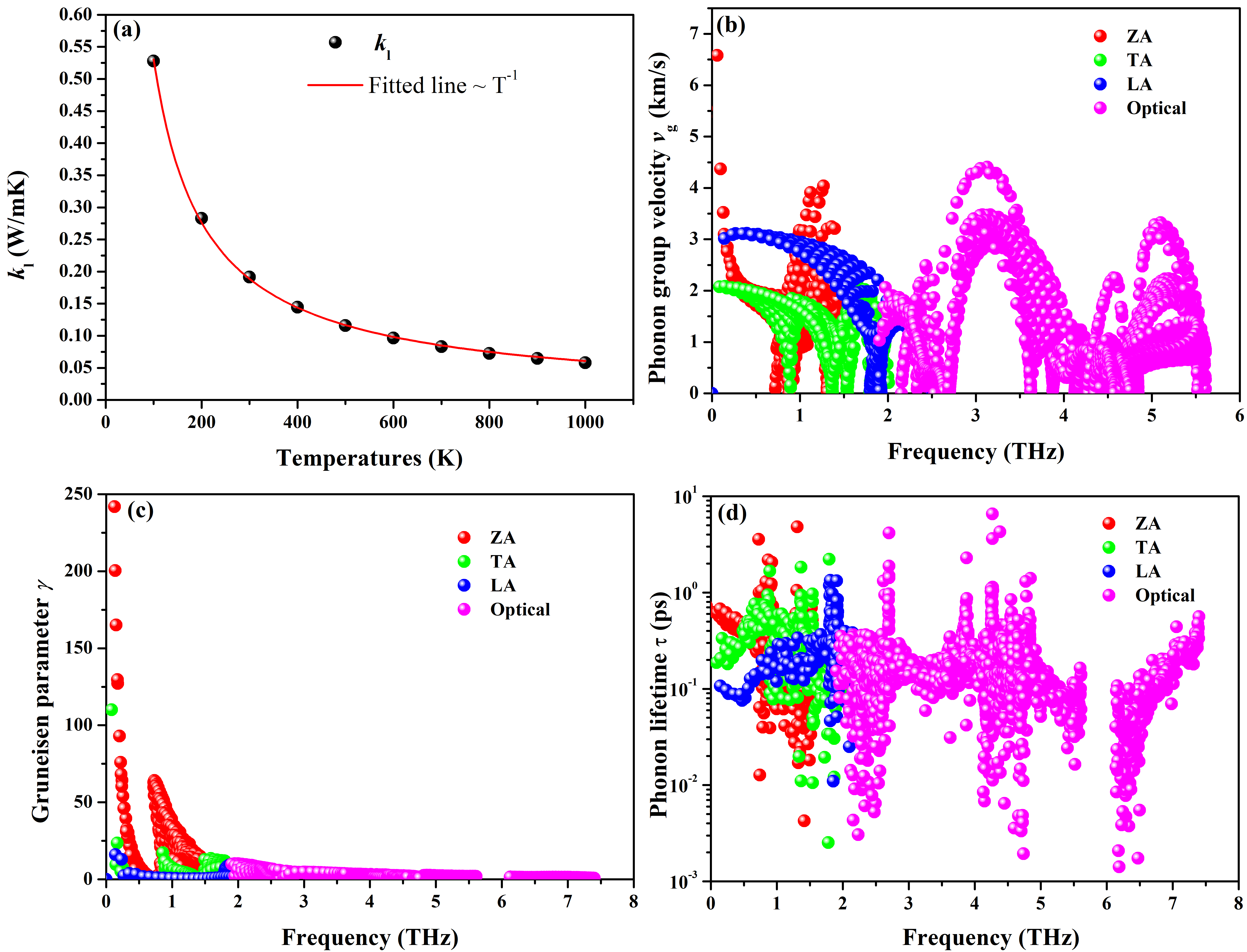}
	\caption{ (Color online) For STe$_{2}$ monolayer, (a) the lattice thermal conductivity $\textit{k}_{l}$ (black sphere) various with temperatures, and the fitted curve (solid red line) conform to the T$^{-1}$ law. (b) shows the phonon group velocity v$_{g}^{2}$ as a function of phonon frequency. (c) represents the Grüneisen parameter as a function of phonon frequency. (d) represents the phonon lifetime as a function of phonon frequency, with different colored spheres representing ZA(red sphere), TA(green sphere), LA(blue sphere), and Optical(magenta sphere) branches, respectively.}
	\label{fig:fig5}
\end{figure}
\begin{figure}
	\centering
	\includegraphics[width=1.0\linewidth]{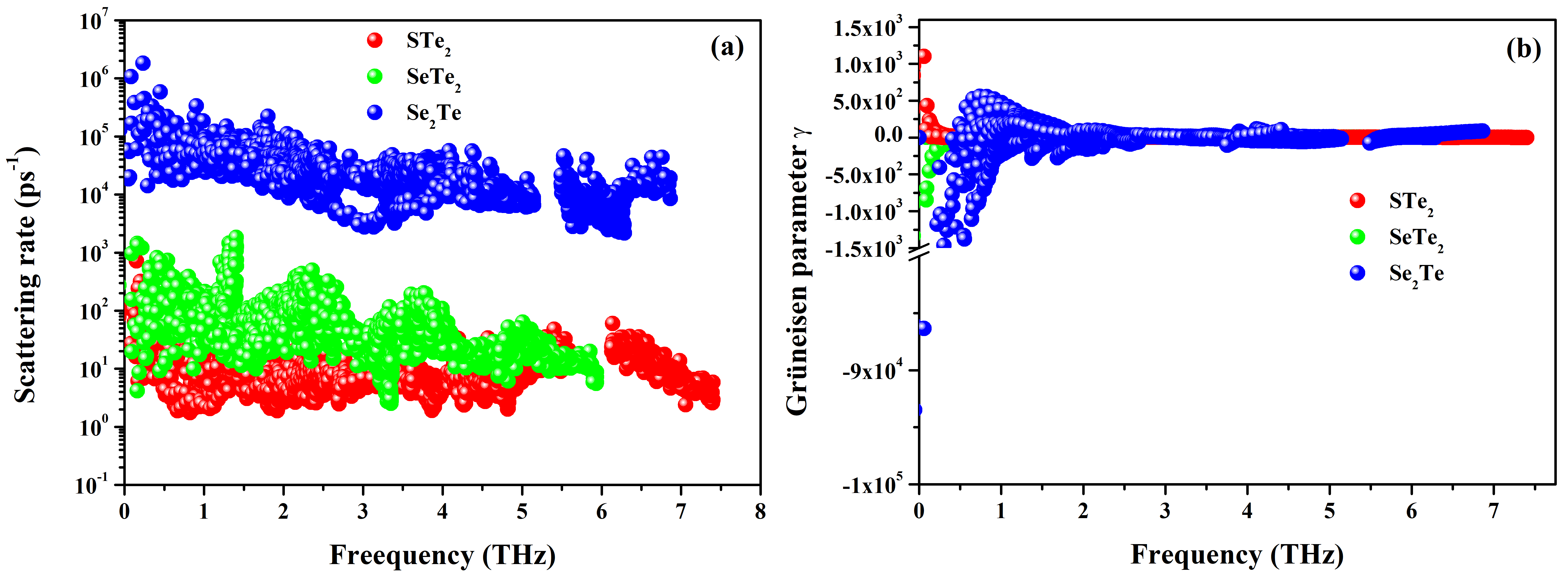}
	\caption{(Color online) (a) Phonon scattering rate and (b) Grüneisen parameters of Janus systems.}
	\label{fig:fig6}
\end{figure}

\begin{figure}
	\centering
	\includegraphics[width=1.0\linewidth]{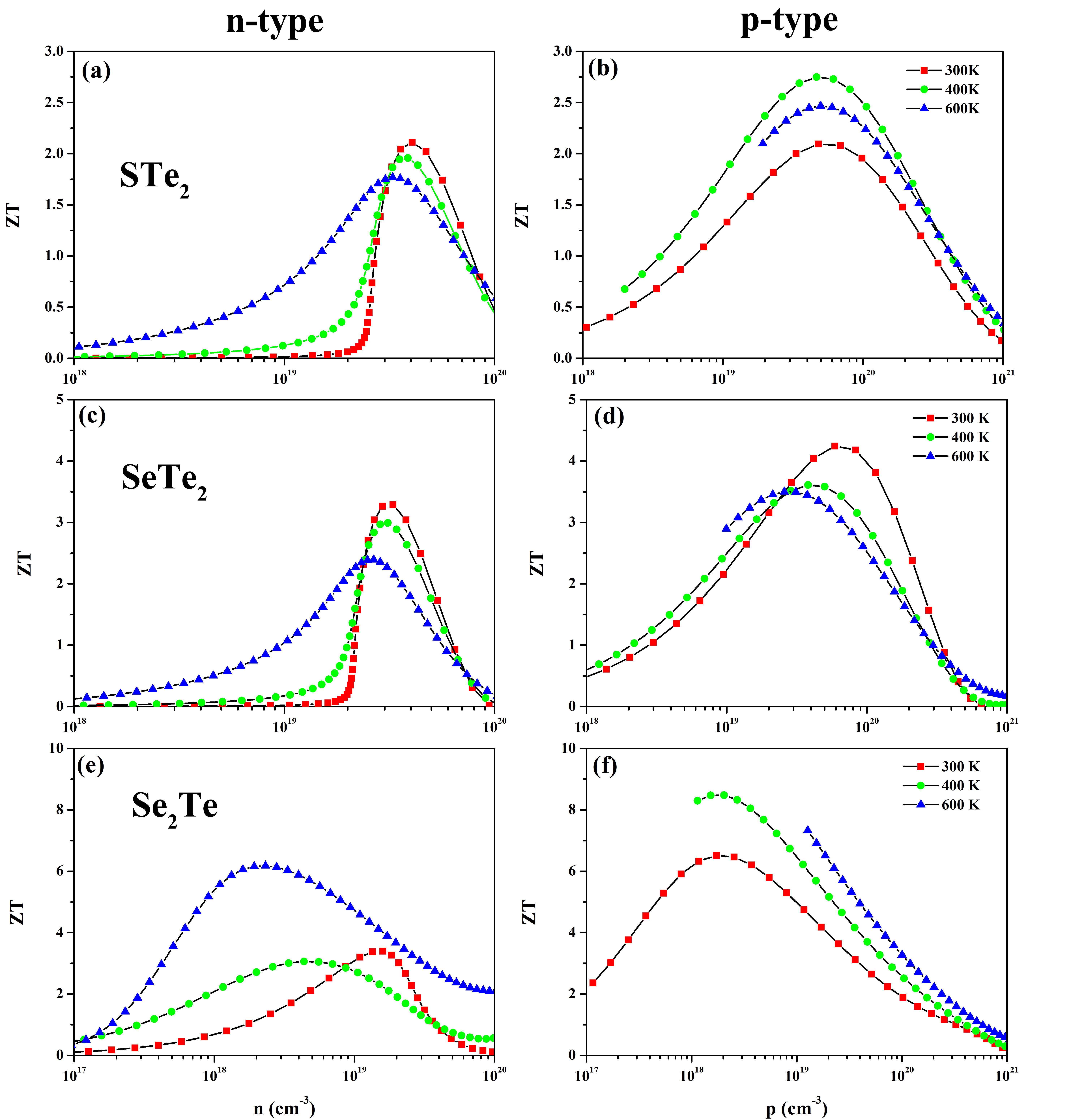}
	\caption{(Color online) The figure of merit \textit{ZT} of (a, b) STe$_{2}$, (c, d) SeTe$_{2}$, and (e, f) Se$_{2}$Te monolayers as a function of doping concentration and temperature.}
	\label{fig:fig7}
\end{figure}

\begin{equation}
RMS\left(\Phi_{i j}\right)=\left[\frac{1}{9} \sum_{a, \beta}\left(\Phi_{i j}^{a \beta}\right)^{2}\right]^{\frac{1}{2}}
\label{Eq3}
\end{equation}
where the $\Phi_{i j}$ represents the second-order IFCs tensor. Large RMS can roughly reveal large interaction strengths \cite{RN2573, RN2571}. The longer the interatomic distance, the smaller the interatomic force and anharmonic interaction \cite{RN2570}. Taking the STe$_{2}$ monolayer as an example, RMS can be calculated by calculating the second-order force tensor, as shown in FIG. \ref{fig:fig4}(a). It can be seen that the value of the RMS tends to decrease with the increase of interatomic distance. Combined with the analysis in FIG. \ref{fig:fig4}(b), it can be found that the cutoff distance of binary Janus STe$_{2}$ monolayer is as high as 8.33 \AA \quad (the 9$^{th}$ nearest neighbor) with strong anharmonic interaction, and the $\textit{k}_{l}$ with good convergence can be obtained, which is very close to that of the 10$^{th}$ nearest neighbor. Therefore, the lattice thermal conductivity of STe$_{2}$ takes the calculation result of the 9$^{th}$-order nearest neighbor, which balances the calculation accuracy and the calculation cost. Using the same method, after testing and analysis, the 7$^{th}$ and 8$^{th}$ nearest neighbors of SeTe$_{2}$ and Se$_{2}$Te can be used to obtain satisfactory converges of lattice thermal conductivity $\textit{k}_{l}$.\\

As summarized in Table \ref{tab2}, when taking the van der Waals radius of the element into account \cite{RN2033}, the corresponding effective thicknesses of Te$_{2}$S, Te$_{2}$Se, and Se$_{2}$Te are 7.12, 7.58, and 7.39 \AA, respectively, in good agreement with other results \cite{RN2560,RN2722}. According to the normalization formula for 2D materials, we obtain the converges lattice thermal conductivities of 0.2, 0.133, and 4.81$\times10^{-4}$ W/mK for STe$_{2}$, SeTe$_{2}$, and Se$_{2}$Te 300 K, respectively(see Table \ref{tab2}). These values are smaller than that of the same family monolayers $\alpha$-Te (9.84 W/mK)\cite{RN2611}, $\alpha$-Se (3.04 W/mK)\cite{RN2612}, square-Te (0.61 W/mK), and square-Se (2.33 W/mK) \cite{RN2613}, illustrating that Janus STe$_{2}$, SeTe$_{2}$, and Se$_{2}$Te monolayers have excellent thermoelectric properties. Combined with Fig. \ref{fig:fig4}(b) and other calculation data, it can be found that the selection of cutoff radius ($n^{th}$-order nearest neighbor) plays a vital role in the accuracy and convergence of lattice thermal conductivity, that is, only by selecting a large enough cutoff radius can the lattice thermal conductivity with good convergence be obtained. Furthermore, the use of different computing software (ShengBTE or phono3py, or AlmaBTE) and computing methods (iteration or RTA) also has a large impact on the results. Ultra-low $\textit{k}_{l}$ value is attributed to the strong coupling effect between the acoustic mode and the low-frequency optical branch (FIG.\ref{fig:fig1}(d, e, f)), which results in strong photoacoustic interactions that suppress phonon transmission.\\
To further insight into the origin of the ultra-low lattice thermal conductivity, the phonon group velocity ($v_{g}$), the Grüneisen parameter ($\gamma$), and the phonon lifetime ($\tau$) of STe$_{2}$ monolayer are plotted in Fig.\ref{fig:fig5}. The average phonon group velocity of Janus $\alpha$-STe$_{2}$ (0.77 km/s) is lower than that of $\alpha$-TeSSe (0.84 km/s) \cite{RN2345}, which is responsible for the lower lattice thermal conductivity of Janus $\alpha$-STe$_{2}$, because $\textit{k}_{l}$ is proportional to the phonon group velocity squared ($v_{g}^{2}$). In addition, the larger $\gamma$ parameter of $\alpha$-STe$_{2}$ means larger anharmonicity resulting from the phonon scattering, which limits phonon transport and thus leads to lower $\textit{k}_{l}$ values. Furthermore, the relatively small phonon lifetime $\tau$ plays a dominant role in the lattice thermal conductivity. Especially, the phonon lifetime of the acoustic branches (ZA, TA, and LA). This also reasonably explains the small value of $\textit{k}_{l}$.  Furthermore, one can see that the Janus Se$_{2}$Te has the lowest lattice thermal conductivity in these three systems, which can deduce from that it possess the largest phonon scattering rate and the Grüneisen parameter, as plotted in Fig. \ref{fig:fig6}. In conclusion, the ultra-low lattice thermal conductivity origin from the i) strong coupling effect between the acoustic mode and the low-frequency optical branch, ii) low phonon group velocity, iii) small phonon lifetime, and iv) large anharmonicity.\\
Finally, combining the calculated thermoelectric power factor \textit{PF}, lattice thermal conductivity $\textit{k}_{l}$, and electron thermal conductivity $\textit{k}_{e}$, we calculate the figure of merit ($\textit{ZT}$) of the three materials at 300, 400 and 600 K temperatures, as shown in Fig.\ref{fig:fig7}. At the same temperature, the $\textit{ZT}$ of the three materials increases first and then decreases with the carrier concentration, regardless of the n-type or p-type carrier type. Considering the effect of temperature, only the n-type $\textit{ZT}$ of STe$_{2}$ and the n(p)-type $\textit{ZT}$ of SeTe$_{2}$ decrease with the increase of temperature, while there is no obvious rule for the other types. This is mainly due to the coupling effect among the Seebeck coefficient, conductivity, electronic thermal conductivity, and lattice thermal conductivity, which check and balance each other and change with the temperature change. Therefore, there is no clear rule to follow between ZT and temperature. As listed in Table  \ref{tab2}, at 300 K, the maximum \textit{ZT} values of n-type (p-type) carrier doping of STe$_{2}$, SeTe$_{2}$ and Se$_{2}$Te are 2.11 (2.09), 3.28 (4.24) and 3.40 (6.51), respectively, which are comparable to or even better than the classical thermoelectric materials SnSe (2.63, 2.46), SnS (1.75, 1.88), GeSe (1.99, 1.73) \cite{RN2618}. By comparison, the calculated results agree with the ones in the reference.  \cite{RN2722}.  Those results show that STe$_{2}$, SeTe$_{2}$, and Se$_{2}$Te are thermoelectric materials with excellent performance and have great potential in thermoelectric device applications.

\section{Conclusion}
\label{sec:Conclusion}
We systematically studied the electrical transport, thermal transport, and thermoelectric properties of Janus STe$_{2}$, SeTe$_{2}$, and Se$_{2}$Te monolayers by first-principles calculations. We illustrate the opposite dependence of the Seebeck coefficient and conductivity on temperature. The decrease of \textit{PF} with temperature is mainly caused by the decrease of $\sigma$. To obtain accurate and convergent lattice thermal conductivity, we calculate RMS to obtain a reasonable cutoff radius for the calculation of third-order forces. The selection of cutoff radius ($n^{th}$-order nearest neighbor) plays a vital role in the accuracy and convergence of lattice thermal conductivity. At room temperature, Janus STe$_{2}$, SeTe$_{2}$, and Se$_{2}$Te monolayers exhibit ultra-low lattice thermal conductivity of 0.2, 0.133, and 4.81$\times10^{-4}$ W/mK, which result from the strong coupling effect between the acoustic mode and the low-frequency optical branch, low phonon group velocity, small phonon lifetime, and large anharmonicity. Hence, high $\textit{ZT}$ values of 2.11 (2.09), 3.28 (4.24), and 3.40 (6.51) for n-type (p-type) carrier doping of STe$_{2}$, SeTe$_{2}$, and Se$_{2}$Te are obtained, indicating that they have important applications in the field of thermoelectric devices. \\

$\textbf{Conflicts of interest:}$
There are no conflicts to declare.

\begin{acknowledgements}
This work was supported by the National Natural Science Foundation of China (Grant No. 12074274), the National Natural Science Foundation of China Academy of Engineering Physics and jointly set up ``NSAF'' joint fund (Grant No. U1830101).
\end{acknowledgements}
\bibliographystyle{apsrev4-1}
\bibliography{Ref}

\end{document}